\documentclass[twocolumn]{aastex631}

\newcommand*\patchAmsMathEnvironmentForLineno[1]{
  \expandafter\let\csname old#1\expandafter\endcsname\csname #1\endcsname
  \expandafter\let\csname oldend#1\expandafter\endcsname\csname end#1\endcsname
  \renewenvironment{#1}
     {\linenomath\csname old#1\endcsname}
     {\csname oldend#1\endcsname¥endlinenomath}}

\accepted{17-Jan-2024}

\shorttitle{Three-Dimensional Velocity Diagnostics of Tycho's SNR}
\shortauthors{Uchida, Kasuga et al.}

\begin{document}

\title{Three-Dimensional Velocity Diagnostics to Constrain the Type Ia Origin of Tycho's Supernova Remnant}


\correspondingauthor{Hiroyuki Uchida}
\email{uchida@cr.scphys.kyoto-u.ac.jp}


\author[0000-0003-1518-2188]{Hiroyuki Uchida}
\affiliation{Department of Physics, Kyoto University, Kitashirakawa Oiwake-cho, Sakyo-ku, Kyoto 606-8502, Japan}

\author[0000-0001-5241-9768]{Tomoaki Kasuga}
\affiliation{Department of Physics, Graduate School of Science, The University of Tokyo, 7-3-1 Hongo, Bunkyo-ku, Tokyo 113-0033, Japan}

\author[0000-0003-2611-7269]{Keiichi Maeda}
\affiliation{Department of Astronomy, Kyoto University, Kitashirakawa, Oiwake-cho, Sakyo-ku, Kyoto 606-8502, Japan}

\author[0000-0002-2899-4241]{Shiu-Hang Lee}
\affiliation{Department of Astronomy, Kyoto University, Kitashirakawa, Oiwake-cho, Sakyo-ku, Kyoto 606-8502, Japan}
\affiliation{Kavli Institute for the Physics and Mathematics of the Universe (WPI), The University of Tokyo, 5-1-5 Kashiwanoha, Kashiwa, Chiba 277-8583, Japan}

\author[0000-0002-4383-0368]{Takaaki Tanaka}
\affiliation{Department of Physics, Konan University, 8-9-1 Okamoto, Higashinada, Kobe, Hyogo 658-8501, Japan}



\author[0000-0003-0890-4920]{Aya Bamba}
\affiliation{Department of Physics, Graduate School of Science, The University of Tokyo, 7-3-1 Hongo, Bunkyo-ku, Tokyo 113-0033, Japan}
\affiliation{Research Center for the Early Universe, School of Science, The University of Tokyo, 7-3-1 Hongo, Bunkyo-ku, Tokyo 113-0033, Japan}
\affiliation{Trans-Scale Quantum Science Institute, The University of Tokyo, 7-3-1 Hongo, Bunkyo-ku, Tokyo 113-0033, Japan}


\begin{abstract}
While various methods have been proposed to disentangle the progenitor system for Type Ia supernovae, their origin is still unclear.
Circumstellar environment is a key to distinguishing between the double-degenerate (DD) and single-degenerate (SD) scenarios since a dense wind cavity is expected only in the case of the SD system.
We perform spatially resolved X-ray spectroscopy of Tycho's supernova remnant (SNR) with XMM-Newton and reveal the three-dimensional velocity structure of the expanding shock-heated ejecta measured from Doppler-broadened lines of intermediate-mass elements. 
Obtained velocity profiles are fairly consistent with those expected from a uniformly expanding ejecta model near the center, whereas we discover a rapid deceleration ($\sim4000$~km~s$^{-1}$ to $\sim1000$~km~s$^{-1}$) near the edge of the remnant in almost every direction.
The result strongly supports the presence of a dense wall entirely surrounding the remnant, which is  confirmed also by our hydrodynamical simulation.
We thus conclude that Tycho's SNR is likely of the SD origin.
Our new method  will be  useful for understanding progenitor systems of Type Ia SNRs in the era of high-angular/energy resolution X-ray astronomy with microcalorimeters.
\end{abstract}


\section{Introduction} \label{sec:intro}
The origin of Type Ia supernovae (SNe) has not yet been clarified. 
While Type Ia SN explosion is known to be caused by a white dwarf (WD) in a binary system, two scenarios are under debate: the double-degenerate \citep[DD;][]{Iben84,Webbink84} or single-degenerate \citep[SD;][]{Whelan73}.
Since only  SD systems require a stellar companion, a detection of the surviving star could be a smoking gun for this scenario.
Many previous optical surveys, however, failed to detect such stars in supernova remnants (SNRs) or pre-explosion data so far \citep[e.g.,][]{Gonzalez12}.
One of the well-observed SNRs in this context is Tycho's SNR (SN~1572).
A promising candidate for a surviving star, namely Tycho~G, was proposed by \citet{Ruiz-Lapuente04}.
Followup observations however suggest that it is unlikely associated with SN~1572 \citep{Kerzendorf09, Xue15}.
Other nearby stars have also been  claimed as alternative candidates, but they are still not decisive \citep{Ihara07, Kerzendorf13, Kerzendorf18}.

Recent theories and observations suggest that the companion stars might be too faint to detect \citep{Hachisu12, Li11}, which leads us to search for another method  to distinguish the two scenarios.
The surrounding environment provides key information; as found by \citet{Zhou16}, expanding molecular bubble exists around Tycho's SNR, which is likely to be evidence for a SD progenitor.
Indirect evidence for the SD scenario also comes from a proper motion study of Tycho's SNR with Chandra \citep{Tanaka21}.
They discovered a drastic deceleration of the blast waves in the southwest during the last $\sim15$~yr observations, suggesting that a dense wall is surrounding a low-density cavity and this environmental structure was formed by a mass-accreting white dwarf before the explosion \citep[see also,][]{Kobashi23}.

If the cavity wall exists around Tycho's SNR and is/was hit by the blast waves in all directions, another piece of supporting evidence may be found in the information along the line of sight.
It is generally known that when an SNR expands into a dense material, a ``reflection (reflected) shock'' begins to move backward behind a contact discontinuity, as is shown by a simple fluid dynamical calculation  \citep[e.g.,][]{Hester94}.
Several X-ray observations of proper motions suggest that backward-moving ejecta or filaments found in Cassiopeia~A \citep{Sato18} and RCW~86 \citep{Suzuki22} are likely reflection shocks produced by ambient molecular clouds or cavity walls.
On the other hand, a line-of-sight velocity structure is, in general, difficult to detect due to a lack of the energy resolution of current X-ray detectors.

In the case of Tycho's SNR,  line-of-sight velocities were partially measured \citep{Sato17a, Millard22} and global expansion structure was reported \citep{Furuzawa09, Hayato10}.
A more comprehensive investigation using Chandra data is recently reported by \citet{Godinaud23}.
In this paper, in order to assess the presence of the cavity wall, we aim to reveal a global velocity structure of the expanding ejecta of Tycho's SNR using the large dataset of XMM-Newton.
Throughout the paper, the distance of Tycho's SNR is  assumed to be 3~kpc \citep[cf.][]{Williams16}.
Errors of parameters are defined as 1$\sigma$ confidence intervals.

\section{Observations} \label{sec:obs}
For the following analysis, we used XMM-Newton data of Tycho's SNR taken for the instrument calibration during 2005--2009  (Table~\ref{tab:obs}).
Note that we did not use Chandra data, while a better angular resolution might be more suitable for our study.
This is because we found that ACIS data are suffering from an effect of ``sacrificial charge'' caused by a bright source \citep[][]{Grant03, Kasuga22phd}, which fills electron traps along readout paths of the CCD and changes charge transfer inefficiency; this effect  results in an overestimation of photon energies.
Among the instruments aboard XMM-Newton, we used MOS data for our analysis because of the better energy resolution than pn.
The event files were processed using the pipeline tool \texttt{emchain} in version 19.1.0 of the XMM-Newton Science Analysis System \citep[SAS;][]{Gabriel04}. 
In all the observations, the whole remnant is within a single CCD (namely CCD1) equipped at the center of the chip array of the MOS.
We used blank-sky data created from background regions of several point-source observations (Table~\ref{tab:obs}); we eliminated each central source region and combined  all the remaining CCD1 data.

\begin{deluxetable*}{cccccc}[th]
\tablenum{1}
\tablecaption{MOS observation log.}
\tablewidth{0pt}
\tablehead{
\colhead{Target Name} & \colhead{Obs.ID} & \colhead{Start Date} & \colhead{Duration (ks)} & \multicolumn{2}{c}{Effective Exp. (ks)} \\
\colhead{} & \colhead{} & \colhead{} & \colhead{}  & \colhead{MOS1} & \colhead{MOS2}
}
\startdata
Tycho's SNR & 0310590101 & 2005.07.03 & 33.4 & 23.4 & 24.1 \\
Tycho's SNR & 0310590201 & 2005.08.05 & 31.9 & 20.1 & 21.6 \\
Tycho's SNR & 0412380101 & 2006.07.28 & 31.9 & 23.0 & 21.4 \\
Tycho's SNR & 0412380201 & 2007.08.21 & 36.9 & 31.5 & 31.9 \\
Tycho's SNR & 0511180101 & 2007.12.30 & 29.1 & 16.7 & 17.3 \\
Tycho's SNR & 0412380301 & 2008.08.08 & 46.5 & 38.2 & 39.1 \\
Tycho's SNR & 0412380401 & 2009.08.14 & 34.3 & 32.8 & 32.8 \\
\hline
Tycho's SNR total &  &  & 244.1 & \multicolumn{2}{c}{374.0}\\
\hline
SWCX-1 & 0402530201 & 2006.06.04 & 96.6 & 83.4 & 85.4 \\
IGR00234+6141 & 0501230201 & 2007.07.10 & 26.9& 24.1 & 24.3 \\
4U0115+63 & 0505280101 & 2007.07.21 & 31.5 & 13.7 & 14.5 \\
4U0241+61 & 0503690101 & 2008.02.28 & 118.4 & 34.6 & 35.2 \\
WD 0127+581 & 0555880101 & 2008.09.03 & 30.9 & 24.5 & 25.6 \\
\hline
Background total &  &  & 304.3 & \multicolumn{2}{c}{365.7}\\
\enddata
\end{deluxetable*}\label{tab:obs}

\section{Analysis and results} \label{sec:ana}

\begin{figure}[!t]
\centering
\includegraphics[width=8.3cm, angle=0]{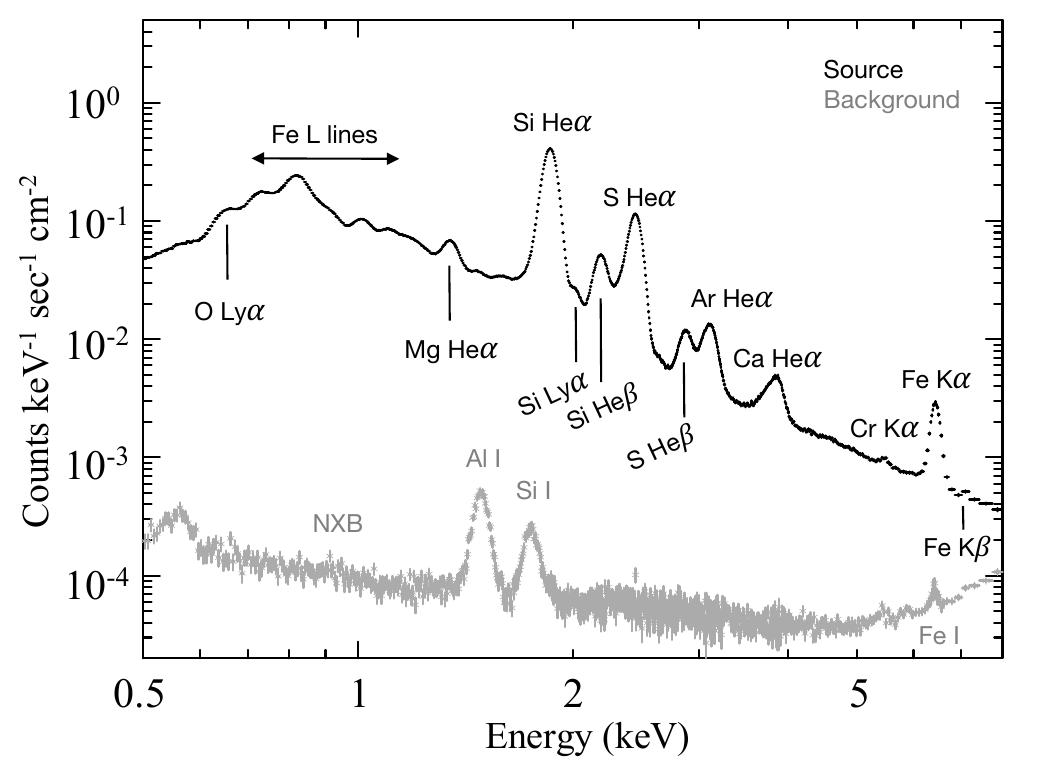}
\caption{X-ray spectrum of Tycho's SNR (black) and  background (grey) obtained with MOS1$+$2.}\label{fig:tycho_spectrum-full}
\end{figure}

Figure~\ref{fig:tycho_spectrum-full} shows the background-subtracted  MOS spectrum of Tycho's SNR.
We detected prominent emission lines of intermediate-mass elements (IMEs; Si, S, Ar, and Ca), and iron-group elements (IGEs; especially Cr and Fe).
Since this remnant is one of the ejecta-dominated SNRs \citep{Miceli15}, these heavy elements including O, Ne, and Mg (hereafter, ONeMg) are all from shock-heated ejecta material.
In Figure~\ref{fig:tycho_spectrum-full}, we also display a non-X-ray background (NXB) spectrum, whose count rate is lower than $\sim1$\% of the SNR flux almost in the entire energy band and  its uncertainty is negligible for our analysis.
The following spectral fit was done on the platform of the X-ray SPECtral fitting package (XSPEC) 12.11.1 \citep{Arnaud96} with the Atomic DataBase (AtomDB) 3.0.9 \citep{Foster15}. 
A fitting method we used was \textit{W}-statistics \citep{Wachter79}, which is a generalized method of \textit{C}-statistics \citep{Cash79} with background subtraction  \citep[see also][]{Kaastra17}.

\begin{figure*}[th]
\gridline{\fig{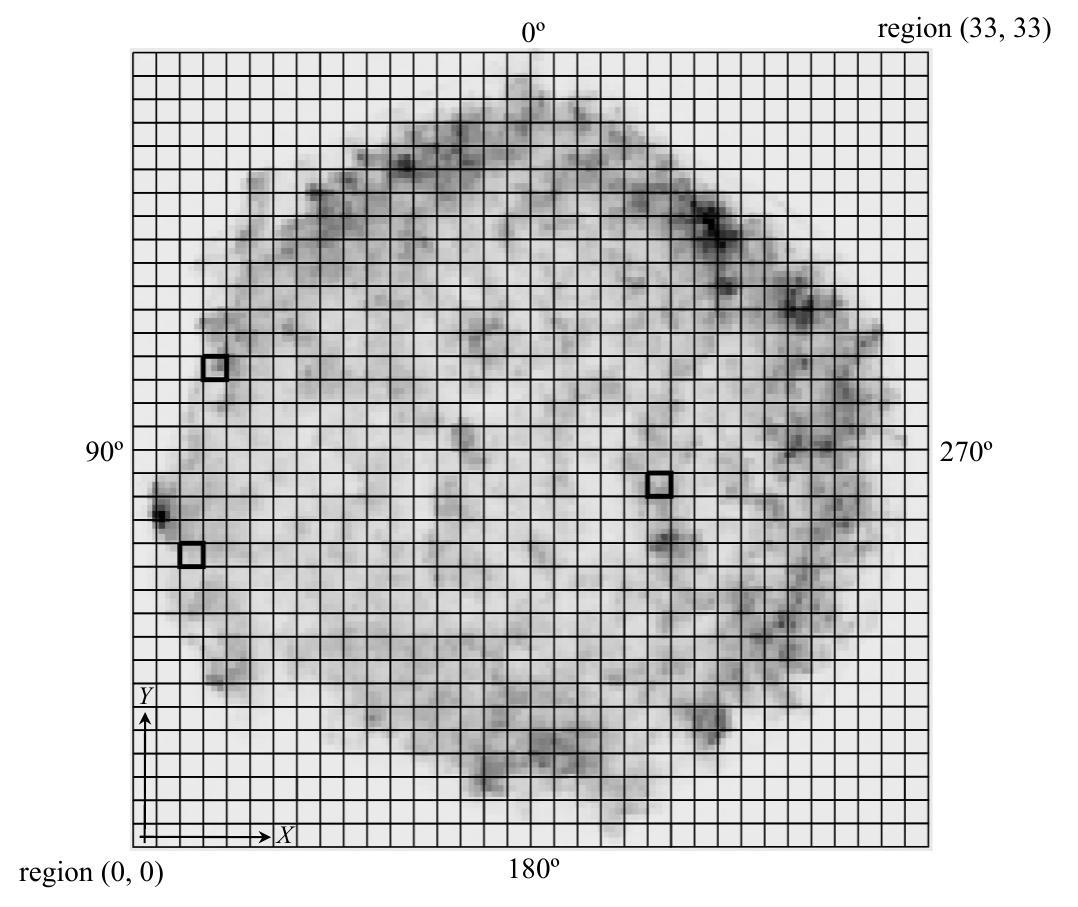}{0.72\textwidth}{}}
\caption{Chandra ACIS image (734~ks in total) of Tycho's SNR (1.80--1.92~keV; Si K$\alpha$ line band) overlaid with the  grids we used for our analysis. The definitions of the region number ($X$, $Y$) and the azimuthal angle $\theta$ (from the north in counterclockwise) are also displayed. The square regions enclosed by the thick lines are those from which we obtained example spectra shown in Figure~\ref{fig:spectrum}.}\label{fig:regions}
\end{figure*}

In order to investigate the spatial distribution of physical properties of Tycho's SNR, we divided the entire region into square grids with a spacing of 15\arcsec \ (totally $34\times34=1156$ segments) as presented in Figure~\ref{fig:regions}: in total 845 segments (grid regions) covering the remnant are available for the spectral analysis.
For each grid region, we performed a wide-band spectroscopy between 0.5~keV and 8~keV. 
Spectral and detector response files were generated with \texttt{mos\_spectra} in SAS.
Although time variations are reported by previous studies with Chandra \citep[cf.][]{Katsuda10,Tanaka21}, a typical expected angular variation is $\sim0.14$\arcsec~yr$^{-1}$, which is much smaller than the grid size and thus negligible for our analysis.
We thus simply combined all the seven observations during 2005--2009 for better statistics for every single grid.
We adopted \texttt{marfrmf}, and \texttt{addrmf}, which are available in the software package of the High Energy Astrophysics Software (HEASoft) 6.28 \citep{Blackburn95}, for combining  ancillary response (arf) and response matrix files (rmf). 

On the basis of the previous X-ray spectroscopic analysis of Tycho's SNR \citep[e.g.,][]{Tamagawa09}, we adopted a multi-component Non-Equilibrium Ionization (NEI) model plus a power-law continuum; the later component is representing the synchrotron emission \citep{Fink94}.
Two Gaussians are also included at $\sim0.7$~keV and $\sim1.2$~keV to compensate for the known uncertainty of the plasma code \citep[see discussion in][]{Okon20}.
The absorption column density $N_{\rm H}$  using the T\"{u}bingen-Boulder model \citep{Wilms00} was fixed at 7.5$\times$10${}^{21}$~cm${}^{-2}$, derived from an average of previous measurements \citep[e.g.,][]{Yamaguchi17,Okuno20}. 

Given that the ejecta consists of pure metals stratified into layers in the shell of the remnant \citep{Hayato10}, we applied several NEI components for a thermal emission corresponding to different element groups, namely ONeMg, IME, and IGE. 
For the ONeMg component, O is fixed to $10^5$ whereas Ne and Mg are free; we confirmed that these abundances do not affect the following analysis and results since the ONeMg component is almost negligible in the middle-energy band that we focus on.
For the IME component, Si is fixed to $10^5$ whereas S, Ar, and Ca are free.
The IGE component is further divided into two components (IGE1 and IGE2) with different plasma temperatures.
For these components, Fe is fixed to $10^5$ and linked the abundances between IGE1 and IGE2.
We tied the electron temperature $kT_{\rm e}$ among the ONeMg, IME and IGE1 components and varied only the ionization timescale $\tau$ for simplicity under an assumption that in each grid the temperature gradient is negligible along the line of sight.
The IGE2 component has a different set of $kT_{\rm e}$ and $\tau$ from the IGE1 component since the ionization state of Fe derived from the line centroid of Fe~K$\alpha$ seems different from that estimated from the Fe-L complex.

The thermal components required for our analysis are then a combination of the low-$kT_{\rm e}$ ONeMg, IME, and IGE1  plus the high-$kT_{\rm e}$ IGE2 (hereafter, multi-NEI component).
We however found that the above model (with a power law) still cannot reproduce spectra well, particularly for regions near the center.
This seems reasonable because the ejecta of Tycho's SNR has a velocity structure as reported by several studies \citep[e.g.,][]{Hayato10, Sato17a}.
We thus applied the two  convolution models (\texttt{vmshift} and \texttt{gsmooth} in XSPEC)  to all the components  for representing a Doppler shift and line broadening effect of each element group (hereafter, we call this ``modified multi-NEI model'').
Note that we set the index parameter in \texttt{gsmooth} to be 1 so that the line width is proportional to the photon energy.

\begin{figure*}[t]
\centering
\includegraphics[width=5.9cm]{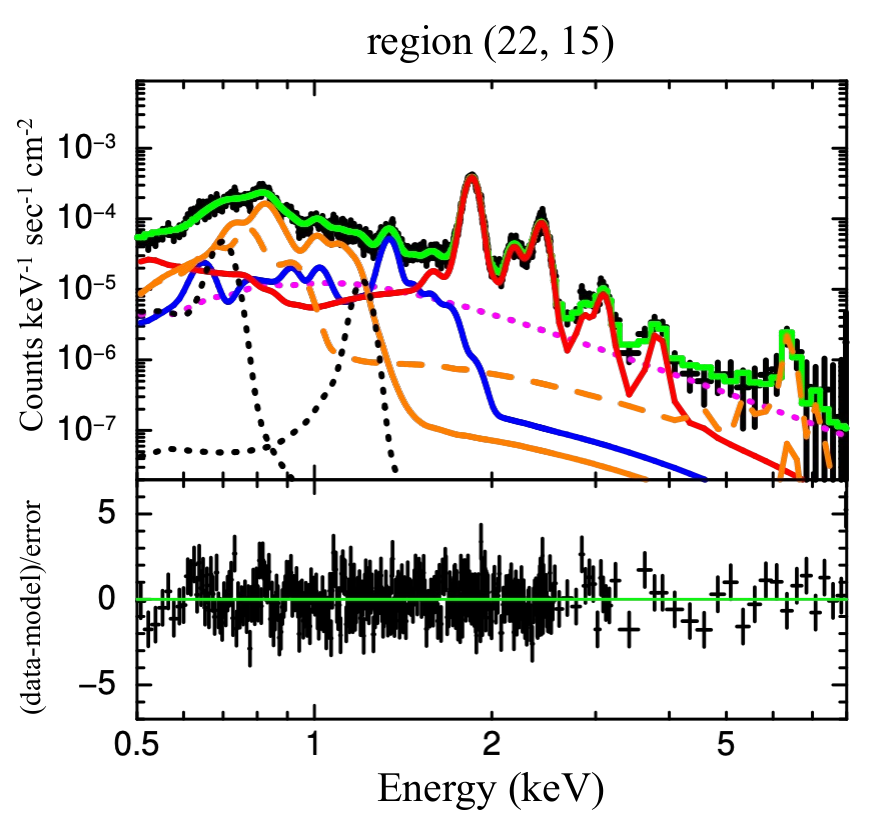}
\includegraphics[width=5.9cm]{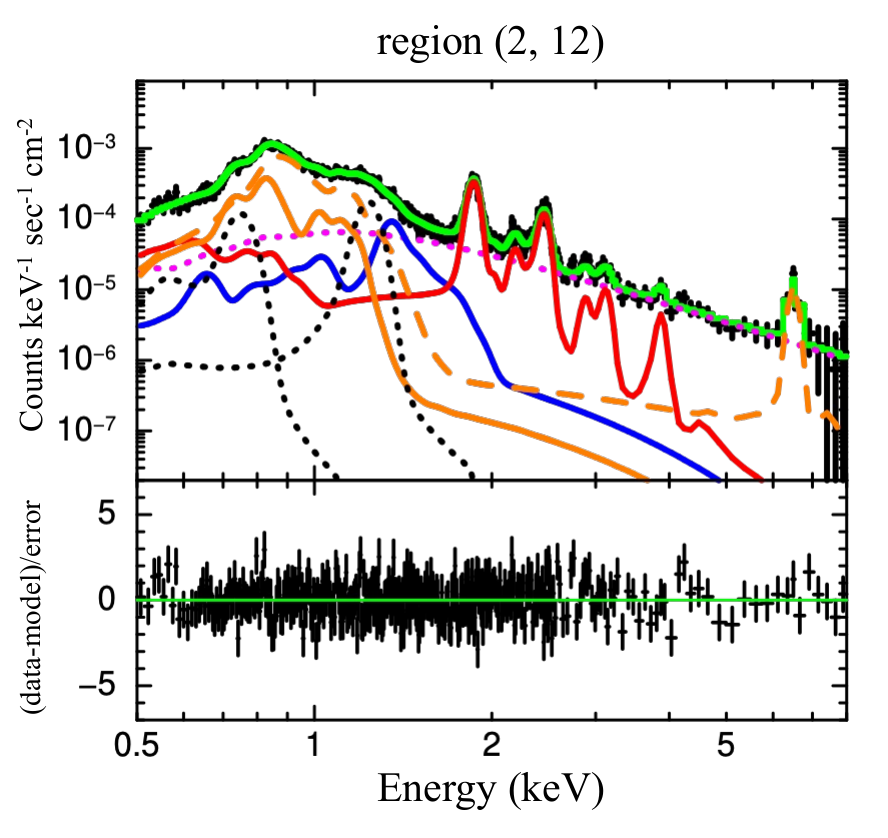}
\includegraphics[width=5.9cm]{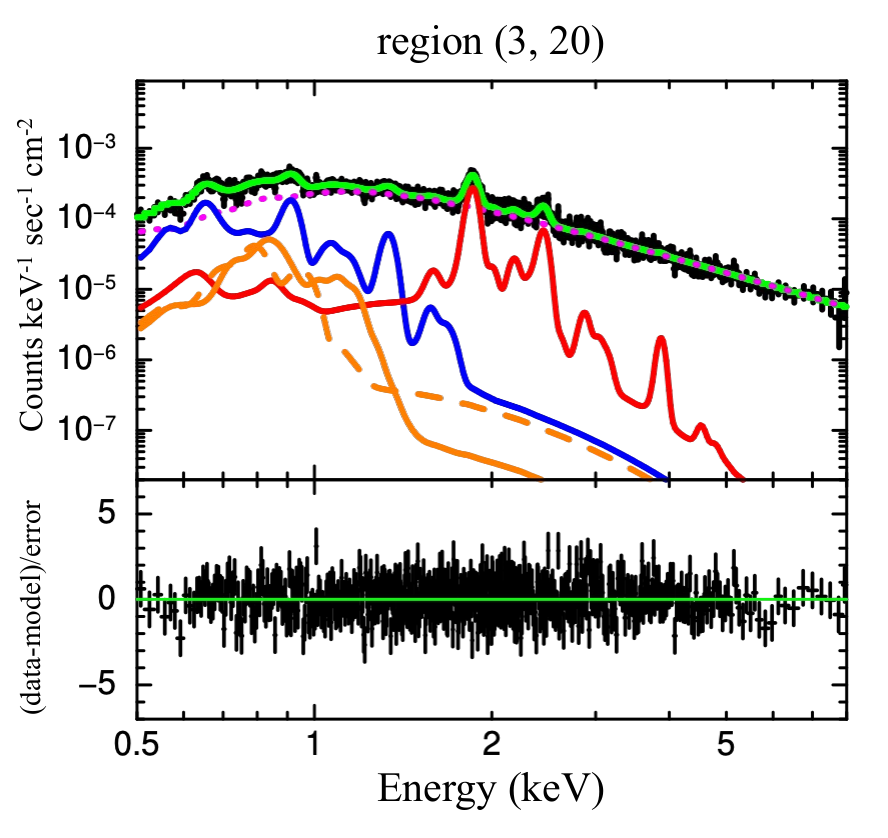}
\caption{X-ray spectra of the example regions shown in Figure~\ref{fig:regions}. The black points are binned observation data and the green solid line is the best-fit model. The blue and red solid lines represent the ONeMg and IME components, respectively. The solid and dashed orange lines indicate the IGE1 and IGE2 components, respectively. The magenta dotted line is the synchrotron emission component. The black dotted lines represent the Gaussians at 0.7~keV and 1.2~keV.
The residuals are displayed in each lower panel.}\label{fig:spectrum}
\end{figure*}




\begin{figure*}[t]
\centering
\includegraphics[width=18cm]{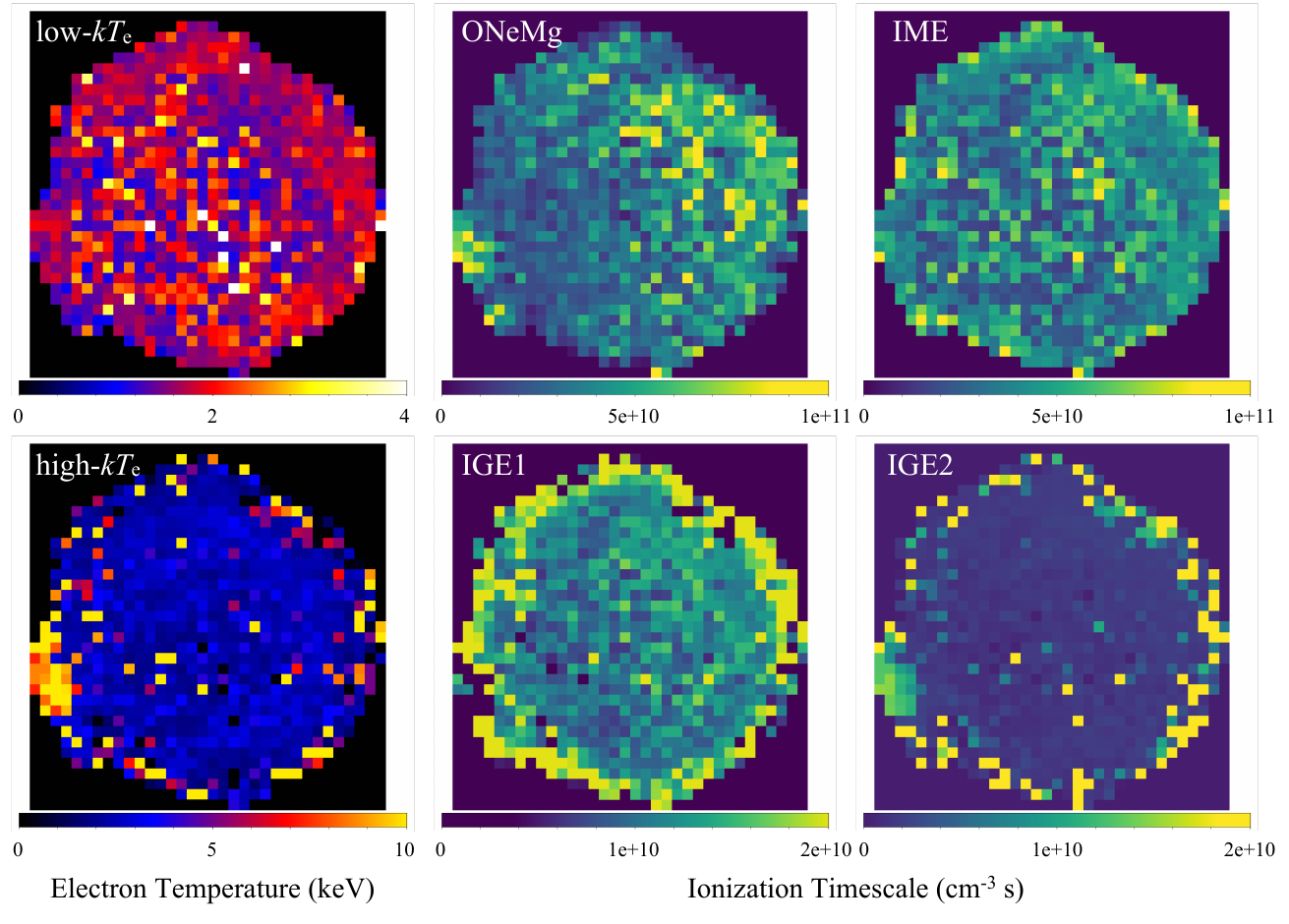}
\caption{Best-fit parameter maps of the electron temperatures $kT_{\rm e}$ and the ionization timescale $\tau$ for each component: low-$kT_{\rm e}$ ONeMg, IME, and IGE1  plus high-$kT_{\rm e}$ IGE2 (see text).}
\label{fig:tycho_nt1}
\end{figure*}

\subsection{Parameter distributions}
In Figure~\ref{fig:spectrum}, we show spectra and best-fit results of example regions.
All the spectra are well reproduced by the modified multi-NEI model.
Parameters of the IGE1 and IGE2 components are especially well constrained even in regions near the outer rim where the nonthermal emission is dominant (for instance, region (3, 20) displayed in Figure~\ref{fig:spectrum}). 
The results enable us to investigate the spatial distributions of the parameters for all the thermal components, as displayed in Figures~\ref{fig:tycho_nt1} and \ref{fig:tycho_map_velocity-full}.

\subsubsection{Electron Temperature and Ionization Timescale}\label{sec:etit}
The electron temperature distributions shown in Figure~\ref{fig:tycho_nt1} indicate that $kT_{\rm e}$ of the ejecta is nearly uniform ($\sim2$~keV) whereas that of the IGE2 (i.e., hot component of Fe) is significantly higher ($>7$~keV) in the southeastern knot \citep[Fe knot;][]{Yamaguchi17} than those in the other regions. 
These results imply that the origin of the IGE2 is different from the low-$kT_{\rm e}$ components, as is pointed out by \citet{Yamaguchi17}.
The overall spatial trend of each component is consistent with those shown by previous studies \citep[e.g.,][]{Miceli15, Matsuda22}.

The ionization timescale $\tau$ for IME shows a relatively uniform distribution, whereas we found that $\tau$ for the ONeMg is somewhat higher in the northwest ($>5\times10^{10}$~cm$^{-3}$~s) than in the other regions ($\sim2\times10^{10}$~cm$^{-3}$~s).
Since the higher-$\tau$ regions roughly coincide with a location of the RS determined by \citet{Warren05}, it might be interpreted as evidence that these elements in the northwest were heated by the revers shock (RS) earlier than in the other directions.
We found that $\tau$ for the IGE1/2 has large uncertainties near the rim.
This is  because the synchrotron emission is dominant in these regions, which makes it difficult to estimate $\tau$ from line intensity ratios.
Note that these uncertainties are not significant for the following discussion because our main purpose is to determine the kinematic structure of the IME component. 


\subsubsection{Three-Dimensional Velocity Diagnostics}\label{sec:dslw}
\begin{figure*}[!ht]
\centering
\includegraphics[width=19cm]{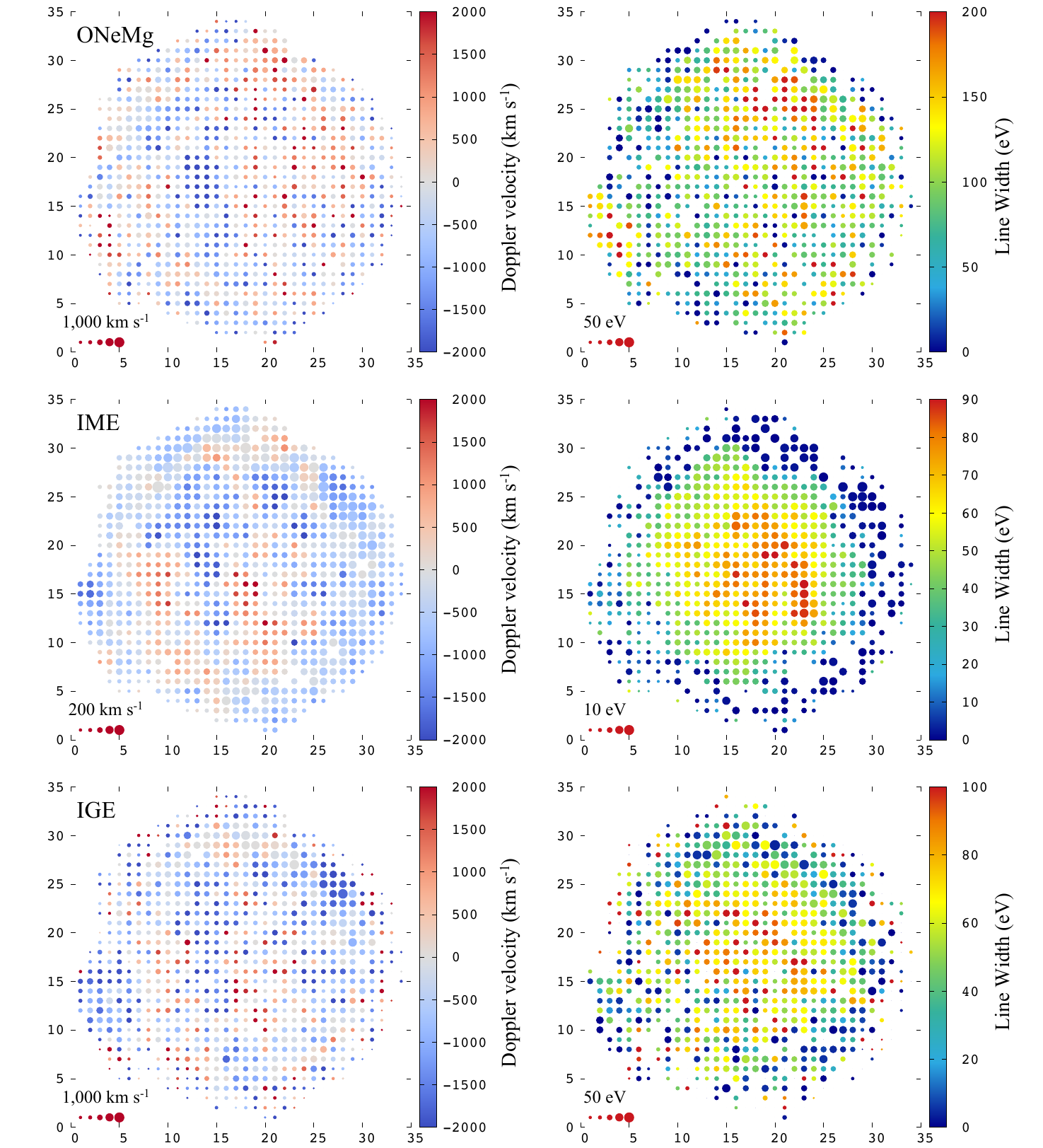}
\caption{Doppler shift (left) and line width (right) maps of Tycho's SNR. Results estimated from the ONeMg, IME, and IGE components are shown in the top, middle, and bottom panels, respectively. Numbers around the maps represent  the $x$$y$ coordinate of  the region  ($X$, $Y$). The size of each dot represents a typical size of the 1$\sigma$ error bars; smaller dots are more uncertain. Reference values of the errors are shown in the bottom-left corner of each panel, where the five dots  from the smallest to largest indicate $\times$3, $\times$2, $\times$1, $\times$1/2, and $\times$1/3 of the reference values. 
The regions where no colored dot is displayed indicate that  the obtained parameter has a large uncertainty and the error exceeds $\times$3 of the reference values.}
\label{fig:tycho_map_velocity-full}
\end{figure*}

Hereafter we primarily focus on the ejecta distribution of the IMEs shown in Figure~\ref{fig:tycho_map_velocity-full}, since systematic uncertainties of the Doppler shift and the line width are  much smaller in comparison with the other components.
The Doppler velocity map of the IMEs shows a clumpy distribution, in which blueshift/redshift regions are somewhat clustering with each other.
The trend is consistent with the mean photon energy maps of Si He$\alpha$ shown by \citet{Sato17a} and \citet{Millard22}.
A recent comprehensive analysis with Chandra also supports our result \citep{Godinaud23}.
The maximum velocity of the IME ejecta is roughly $\sim\pm2,000$~km~s$^{-1}$, which is consistent with the previous measurements for the Si ejecta.
The result is understandable because in the IME component the most prominent line is Si~He$\alpha$.

The Doppler velocity map for the IGEs is weakly correlated with that for the IME: the northwestern and southeastern parts (around the Fe knot) are blueshifted, whereas several clumps in the south are redshifted.
The correlation implies that the global ejecta structure for the IMEs spatially overlaps with that for the IGEs, especially Fe.
On the other hand, the ONeMg component shows no clear correlation with the heavier elements.
This may confirm that these light elements spatially distribute above the core elements such as Fe and Si as  suggested also by the  ionization timescale distribution of the ONeMg component (see Section~\ref{sec:etit}).

\begin{figure*}[!ht]
\centering
\includegraphics[width=18cm]{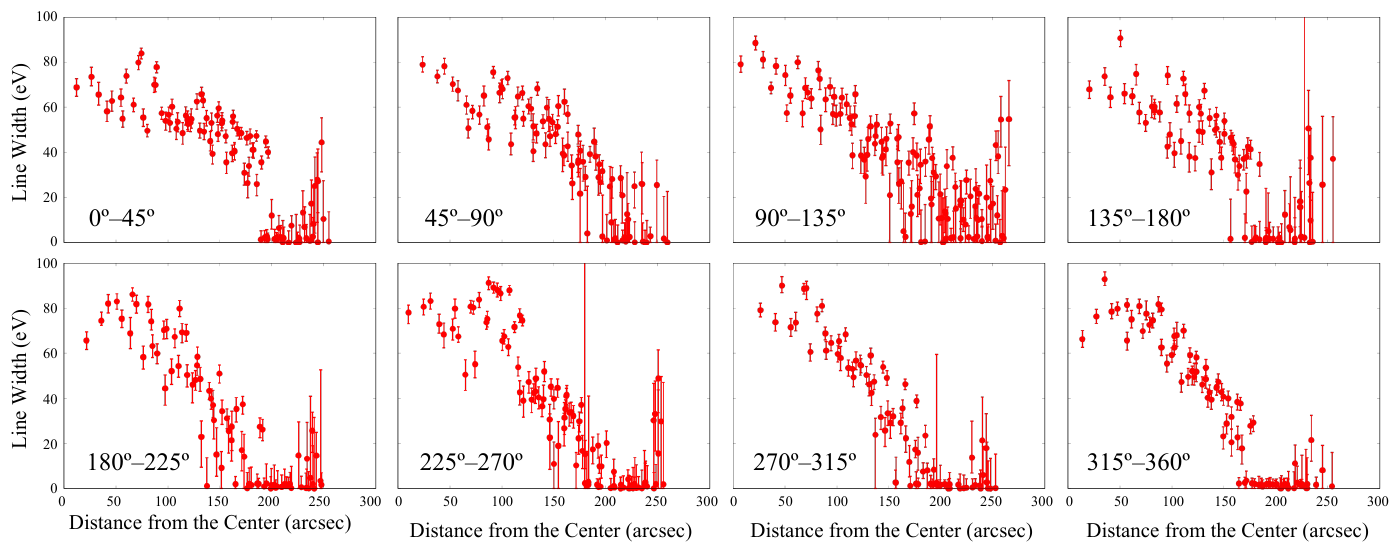}
\caption{Radial profiles of the line width from the center to the outside for each octant. The x-axis corresponds to the projection position $x$ defined in the text and Figure~\ref{fig:tycho_schematics}}
\label{fig:width_profile}
\end{figure*}

From the line width maps, which are the main indicators of the ejecta kinematics, we clearly see that the IMEs show an azimuthally symmetric  velocity structure.
A similar trend is apparent in the line width map of the IGEs whereas that of the ONeMg shows a different asymmetric distribution.
Since the lighter elements are distributed in outer regions, this component is likely  affected by an inhomogeneous ambient density structure.
The heavier elements, especially the IMEs, are expanding symmetrically from a single point (probably from the center of the remnant), as previously reported \citep{Furuzawa09, Hayato10, Sato17a}.
The result will be helpful for constraining the 3D expanding structure of Tycho's SNR with a line-of-sight velocity of the ejecta.

We  constructed azimuthally averaged radial profiles of the obtained line widths for each octant, as displayed in Figure~\ref{fig:width_profile}.
The resultant profiles are similar: line widths ($\sigma$) are highest near the center and gradually decrease toward the rim of the remnant, which is reasonable given that Tycho's SNR is approximately spherically symmetric even along the ling of sight.
Only the profile for $\theta=90^{\circ}$--$135^{\circ}$  deviates from this trend, which can be partially  attributed to thermal broadening in the southeastern knot reported by \citet{Williams20}.
We note that thermal broadening is also expected from the contact discontinuity to the reverse shock front.
According to \citet{Badenes06}, this effect is significant only in a narrow region (less than 10\% of the line-of-sight volume) and is negligible beyond 200~arcsec from the center.

From Figure~\ref{fig:width_profile},  we  found that $\sigma$  falls to $\sim0$~eV far inside the shell.
This trend is clearly exemplified in the profiles for $\theta=180^{\circ}$--$225^{\circ}$ and $315^{\circ}$--360$^{\circ}$.
It may be somewhat counterintuitive because if we assume a normal radial velocity variation of expanding material, $\sigma$ should reach 0~eV  at the outer edge of the profile.
As reported by \citet{Godinaud23}, proper motion velocities ($V_{xy}$) near the rim of the remnant exceed $\sim3,000$--4,000~km~s$^{-1}$.
On the other hand, the line width $\sigma$ we measured represents an average of the superposition of the velocity components in the line-of-sight direction.
These results may be consistently interpreted by assuming that the ejecta has a double-shell velocity structure: an inner thick high-velocity component overlaid with a thin lower-velocity shell.
This assumption is probably reasonable because Chandra HETG observations \citep{Millard22} suggest  lower line-of-sight velocities near the shell.
Given that the trend is common across all the octants, we infer that our result may reflect a global velocity structure of Tycho's SNR.




\begin{figure}[ht]
\centering
\includegraphics[width=7.2cm]{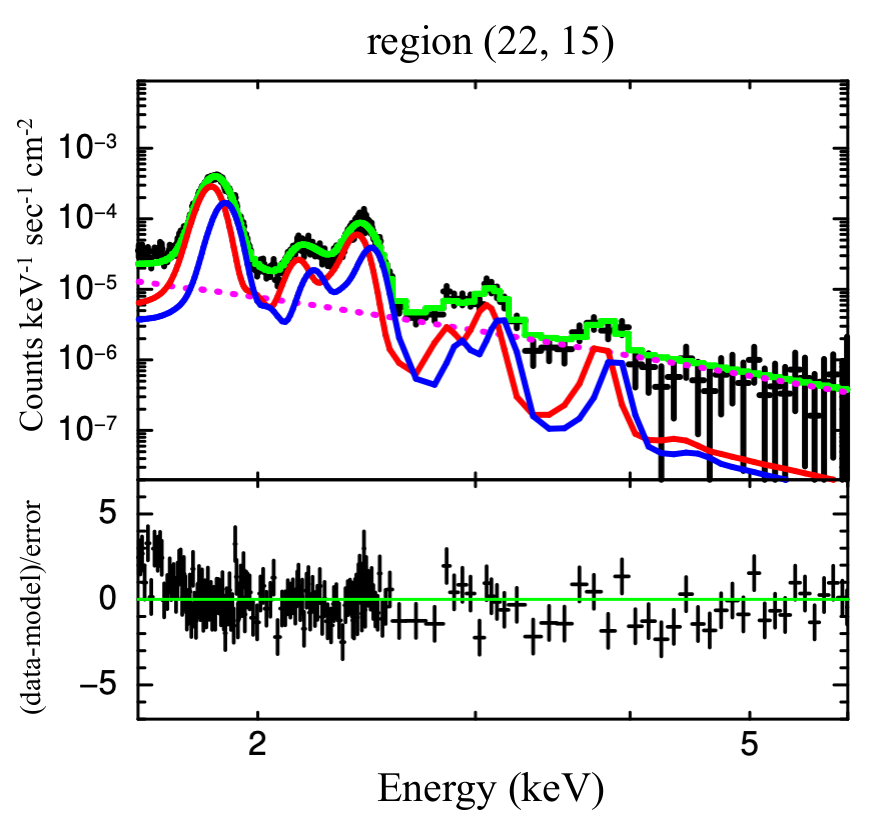}
\caption{Same as Figure~\ref{fig:spectrum}, but with an example best-fit result of the Doppler-broadening model. The red and blue curves indicate the red- and blue-shifted IME component, respectively. }
\label{fig:tycho_spectrum-example2}
\end{figure}

\begin{figure*}[ht]
\centering
\includegraphics[width=18cm]{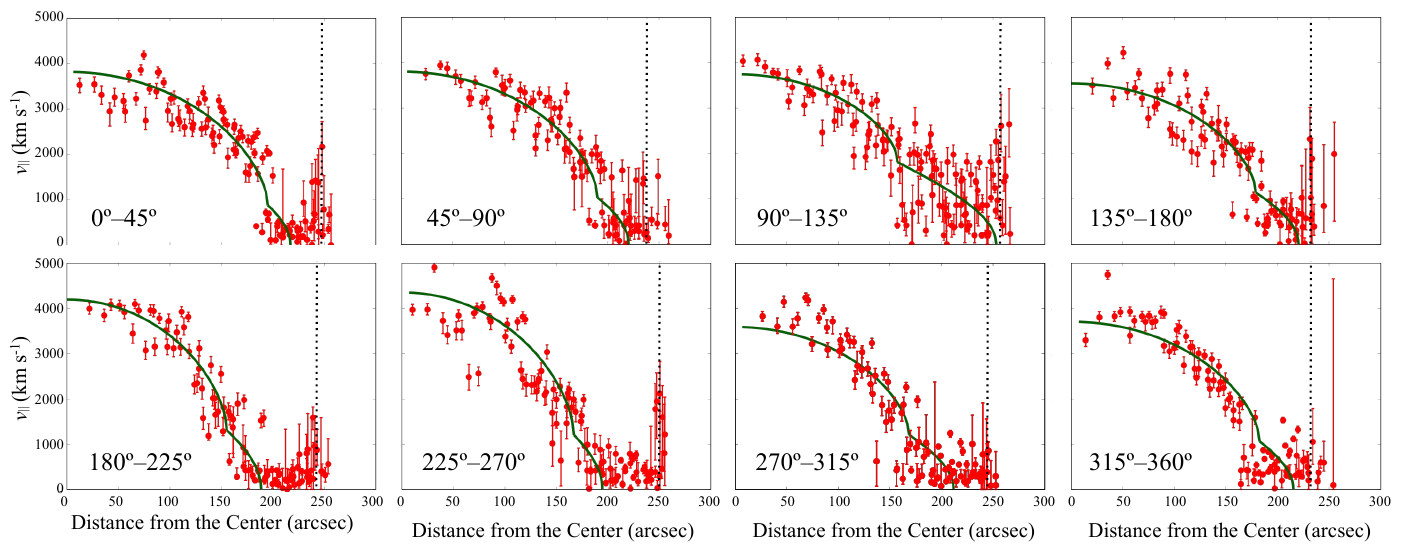}
\caption{Radial profiles of $v_{\parallel}$ from the center to the outside for each octant. Best-fit results of equation~(\ref{eq:v_para}) is shown in green. The vertical dotted lines indicate positions of the contact discontinuity $r_{\rm{CD}}$. }
\label{fig:v_parallel_profile}
\end{figure*}

\section{Discussion} \label{sec:dis}
\subsection{Measurement of the ejecta velocity}\label{sec:ejecta_velocity}
As claimed in Section~\ref{sec:dslw}, the result shown in Figure~\ref{fig:width_profile} implies that some deceleration may have occurred near the edge of Tycho's SNR.
This scenario is possible if we consider a recent interaction between  the blast waves and ambient dense gas.
Since previous studies \citep[e.g.,][]{Zhou16, Tanaka21} suggest the presence of such dense material around the remnant, it is reasonable to consider that the forward shock recently hit the cavity wall essentially in all the directions, which can affect the global radial velocity structure. 

In order to clarify the global structure of Tycho's SNR, we measured the line-of-sight expansion velocity $v_{\parallel}$ of the IME in each segment by refitting the spectra with a Doppler-broadening model, with reference to \citet{Hayato10}.
We focused on the data between 1.6~keV and 6.0~keV because the IME component is generally dominant in the middle-energy band.
For simplicity, we assume that the observed line broadening can be explained by a two-NEI (plus a power law) model with different velocities  $v_{\rm{r}}=v_{\parallel}$ and $v_{\rm{b}}=-v_{\parallel}$, representing the red/blue-shifted (i.e., moving backward and forward) components, respectively.
The parameters of the two NEI components, except for the normalization, were linked to each other and fixed at the best-fit values of the multi-NEI model. 
We also included a common offset parameterized  as $v_{\rm{0}}$ for representing a bulk motion of Tycho's SNR and/or a systematic uncertainty in the instruments.
By averaging the line shift at the outermost edges of the remnant, we obtained $v_{\rm{0}}=-420$~km~sec$^{-1}$.
A local red/blue shift is in this case explained by an intensity ratio of  $v_{\rm{r}}$ and $v_{\rm{b}}$ components.
We confirmed that all the spectra were well fitted with the above model.
An example fit is shown in Figure~\ref{fig:tycho_spectrum-example2}.

\begin{figure}[t]
\centering
\includegraphics[width=8.8cm]{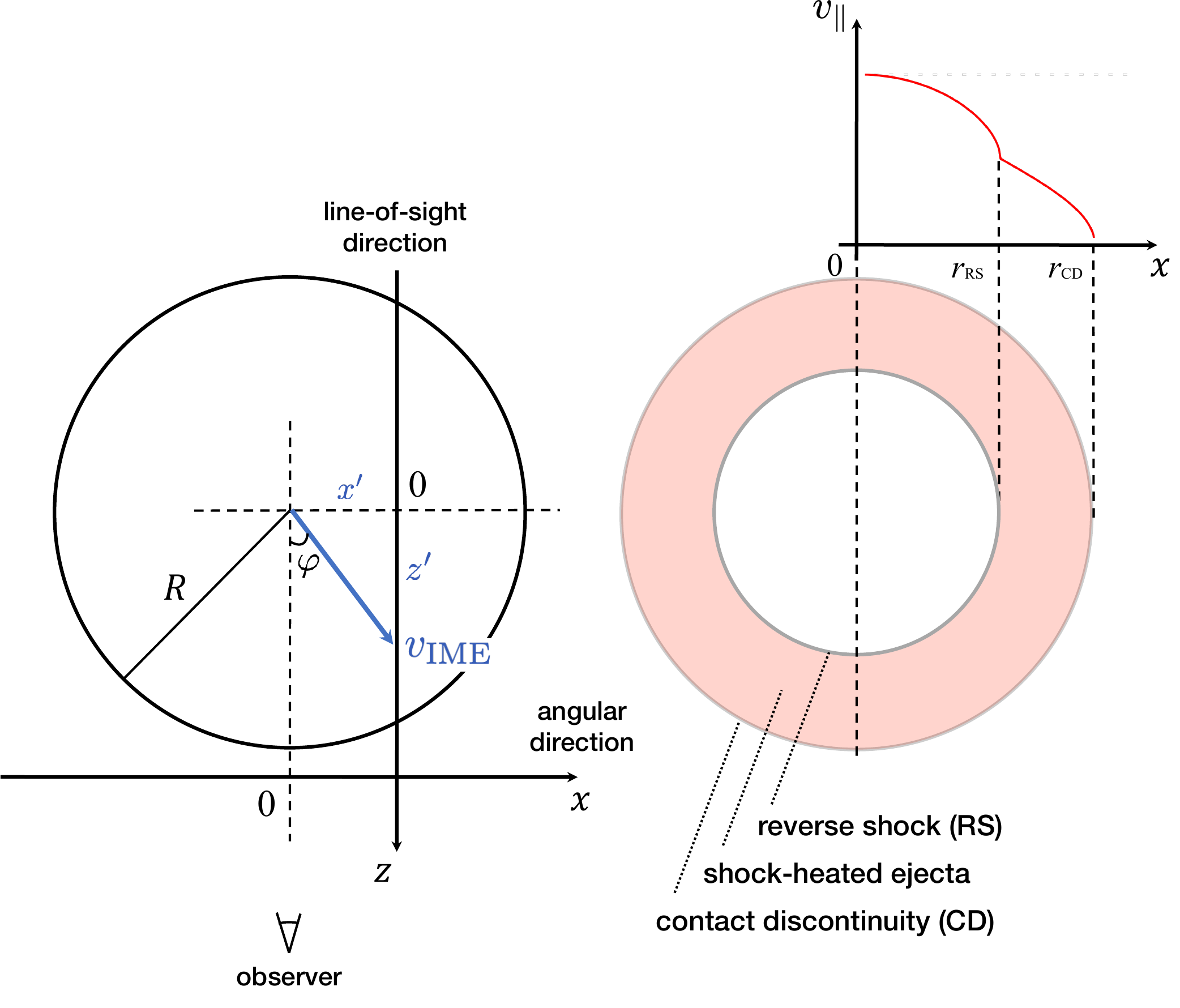}
\caption{Schematic drawing of expanding shock-heated ejecta. Definitions of the parameters are written in the main text and appendix.}
\label{fig:tycho_schematics}
\end{figure}

Figure~\ref{fig:v_parallel_profile} displays the $v_{\parallel}$ profiles for different directions based on the best-fit results with the Doppler-broadening model.
Since $v_{\parallel}$ is measured independently from $v_{\rm{0}}$ and directly reflects the expansion structure, the uncertainty in the latter does not critically affect the following discussion.
Incidentally we found that a similar analysis with Chandra gives a much broader line shape and a stronger blue shift in almost all the regions. 
A possible cause of this discrepancy is the effect of ``sacrificial charge'' mentioned in Section~\ref{sec:obs}.
While it is out of the scope of this paper to argue  this overestimation and calibration issue concerning Chandra, we exemplify spectral comparison between two instruments in Apendix~\ref{sec:app_chandra} that will be useful for future studies.


As indicated in Figure~\ref{fig:v_parallel_profile}, the profile shapes of $v_{\parallel}$ are similar to those of the line widths (Figure~\ref{fig:width_profile}).
We then constructed a model profile for $v_{\parallel}$ under the following assumption.
Since the reverse shock of this remnant has not reached the center yet, the X-ray emitting region is within $r_{\rm{RS}}\le r<r_{\rm{ejecta}}$, where $r_{\rm{RS}}$ and $r_{\rm{ejecta}}$ are the positions of the reverse shock and the outer boundary of the ejecta, respectively.
Assuming a uniform and low-density environment around Tycho's SNR, the expansion velocity of the ejecta (hereafter, $v_{\rm IME}$) within this region can be considered as roughly being constant with radius \citep[e.g.,][]{Blondin01}.
As explained in Appendix~\ref{sec:app} and Figure~\ref{fig:tycho_schematics}, we can describe  $v_{\parallel}(x')$, which is  an average velocity along the line-of-sight at a position $x=x'$, as follows:
\begin{equation}\label{eq:v_para}
v_{\parallel}(x') = \\
\left\{
\begin{array}{ll}
\alpha(x')v_{\rm IME}+\beta(x')v_{\rm IME}  & (0 \leq x' < r_{\rm{RS}}),  \\
 \gamma(x')v_{\rm IME}  & (r_{\rm{RS}} \leq x' < r_{\rm{ejecta}}),\\
0 & (r_{\rm{ejecta}} \leq x'),\\
\end{array}
\right.
\end{equation}
where,
\begin{eqnarray}
\alpha(x') &=&  \frac{r_{\rm{RS}}}{r_{\rm{RS}}+r_{\rm{ejecta}}} \sqrt{1 - \left( \frac{x'}{r_{\rm{RS}}} \right)^2} \\
\beta(x') &=&  \frac{r_{\rm{ejecta}}}{r_{\rm{RS}}+r_{\rm{ejecta}}} \sqrt{1 - \left( \frac{x'}{r_{\rm{ejecta}}} \right)^2} \\
\gamma(x') &=& \sqrt{ \frac{r_{\rm{ejecta}}-x'}{r_{\rm{ejecta}}+x'}}.
\end{eqnarray}



We fitted  the model $v_{\parallel}(x')$ described by equation~(\ref{eq:v_para}) to the $v_{\parallel}$ profiles obtained by the Doppler model applied to the XMM-Newton data as shown in Figure~\ref{fig:v_parallel_profile}.
From the best-fit results, we  found a significant discrepancy between the model and the data
 except for the southeast direction ($\theta=90^{\circ}$--$135^{\circ}, 135^{\circ}$--$180^{\circ}$) where the Fe-knot is located \citep{Yamaguchi17}.
In Figure~\ref{fig:v_parallel_profile}, we also show the locations of the contact discontinuity $r_{\rm{CD}}$, which is estimated as the average radius of the outer boundary of the ejecta in each sector.
If the ejecta is expanding in a uniform and low-density environment without deceleration, $r_{\rm{ejecta}}$ should  coincide with the contact discontinuity ($r_{\rm{CD}}$), i.e., $r_{\rm{ejecta}}=r_{\rm{CD}}$.
This assumption, however, fails to reproduce the observed $v_{\parallel}$ profiles, leading us to favor a non-uniform expanding structure of the ejecta.


\begin{deluxetable}{ccccc}
\tablenum{2}
\tablecaption{Best-fit parameters of $v_{\rm ejeccta}$, $r_{\rm{RS}}$, and $r_{\rm{ejecta}}$ obtained from equation~\ref{eq:v_para} and estimated $r_{\rm{CD}}$.}
\tablewidth{0pt}
\tablehead{
\colhead{Sector} & \colhead{$v_{\rm IME}$ (km~s$^{-1}$)} & \colhead{$r_{\rm{RS}}$}   & \colhead{$r_{\rm{ejecta}}$} & \colhead{$r_{\rm{CD}}$}
}
\startdata
$0^{\circ}$--$45^{\circ}$  & $3730\pm80$ & $197\arcsec\pm5\arcsec$  & $217\arcsec\pm4\arcsec$ & 247\arcsec\\
$45^{\circ}$--$90^{\circ}$  & $3850\pm70$ & $188\arcsec\pm5\arcsec$  & $217\arcsec\pm4\arcsec$&236\arcsec\\
$90^{\circ}$--$135^{\circ}$  & $3810\pm90$ & $157\arcsec\pm10\arcsec$  & $240\arcsec\pm4\arcsec$& 256\arcsec\\
$135^{\circ}$--$180^{\circ}$  & $3630\pm90$ & $177\arcsec\pm6\arcsec$  & $208\arcsec\pm3\arcsec$&233\arcsec \\
$180^{\circ}$--$225^{\circ}$  & $4190\pm120$ & $153\arcsec\pm7\arcsec$  & $189\arcsec\pm4\arcsec$&242\arcsec \\
$225^{\circ}$--$270^{\circ}$  & $4370\pm120$ & $197\arcsec\pm8\arcsec$  & $193\arcsec\pm5\arcsec$&249\arcsec \\
$270^{\circ}$--$315^{\circ}$  &  $3600\pm130$  & $167\arcsec\pm9\arcsec$ & $212\arcsec\pm4\arcsec$&244\arcsec \\
$315^{\circ}$--$360^{\circ}$ & $3740\pm150$ & $167\arcsec\pm11\arcsec$ & $205\arcsec\pm5\arcsec$&232\arcsec \\
\hline
average & $3870\pm40$ & $175\arcsec\pm3\arcsec$ & $210\arcsec\pm1\arcsec$ &242\arcsec \\
\enddata
\end{deluxetable}\label{tab:vrr}

\subsection{Velocity Structure of Tycho's SNR}\label{sec:ejecta_velocity}
From the best-fit $v_{\parallel}$ profiles (Figure~\ref{fig:v_parallel_profile}), we obtained  $v_{\rm IME}$, $r_{\rm{RS}}$, and $r_{\rm{ejecta}}$ as the fitting parameters, for each direction as summarized in Table~\ref{tab:vrr}.
The averaged ejecta velocity $v_{\rm IME}\sim3900$~km~s$^{-1}$ is between the predicted value \citep[$\sim2000$~km~s$^{-1}$;][]{Badenes06} and the past measurement with Suzaku \citep[$4700\pm100$~km~s$^{-1}$;][]{Hayato10}, and in  good agreement with the previous proper-motion measurement with Chandra \citep{Katsuda10}; 0\farcs21--0\farcs31~yr$^{-1}$ ($3000$--4400~km~s$^{-1}$ for a distance of 3~kpc).
We  also found that $v_{\rm IME}$ at  $\theta=180^{\circ}$--$270^{\circ}$ (southwest region) is larger than those at the other directions.
It is roughly aligned with previous studies suggesting encountering dense gas toward the northeast direction in past \citep{Lee04, Zhou16, Williams16, Godinaud23}, and  more recent  deceleration of blast waves in the southwest \citep{Tanaka21}.
The average position of the reverse shock $r_{\rm{RS}}=175\arcsec\pm3\arcsec$ is also fairly consistent with the previous estimations, i.e., 183\arcsec \citep{Warren05} or 158\arcsec \citep{Yamaguchi14a}.
These results indicate that the velocity structure and the ejecta morphology are naturally explained by the standard SNR evolution model \citep[e.g.,][]{Truelove99} except for the outer layers.

\begin{figure*}[ht]
\centering
\includegraphics[width=18cm]{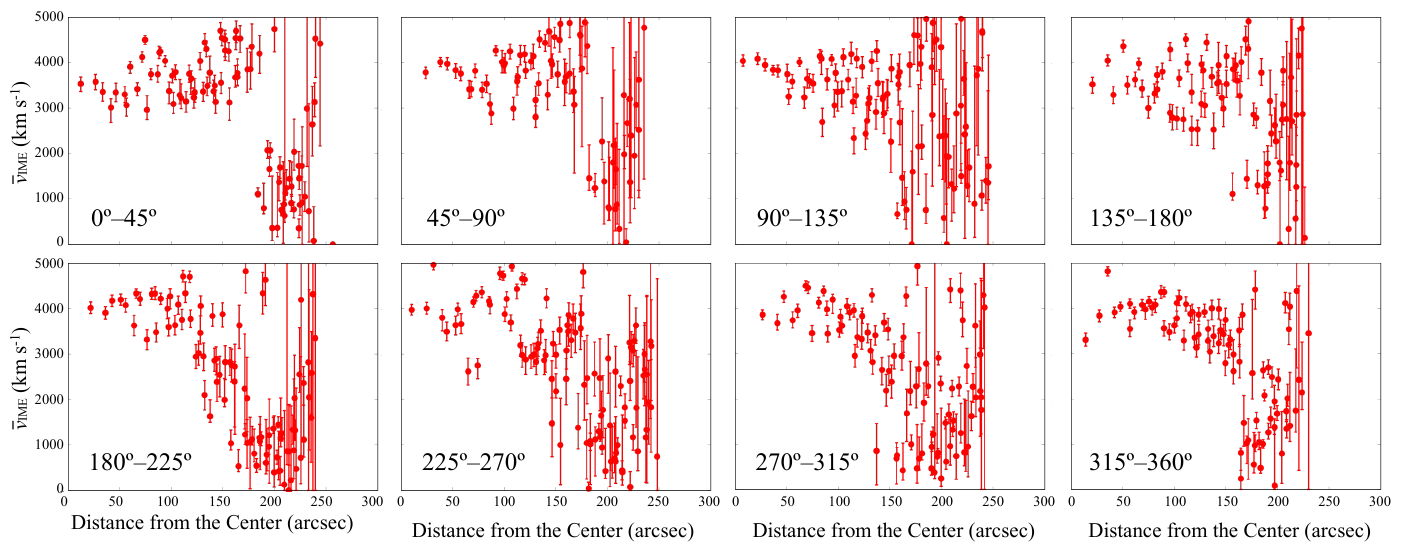}
\caption{Radial profiles of $\overline{v}_{\rm IME}$ for each octant.}
\label{fig:v_IME}
\end{figure*}


In order to understand the cause of  the discrepancy between the $v_{\parallel}(x)$ model and the $v_{\parallel}$ profiles near the rim as seen in Figure~\ref{fig:v_parallel_profile}, we converted the measured $v_{\parallel}$  to $\overline{v}_{\rm IME}$ by solving an inverse problem of  equation~(\ref{eq:v_para}).
Here, we define $\overline{v}_{\rm IME}$ as an ``expected'' averaged velocity of the ejecta along the line of sight.
Using equation~(\ref{eq:v_para}), if the ejecta is not decelerated and has a constant velocity all over the emitting region, the value $\overline{v}_{\rm IME}$ is \textit{expected} to be  constant throughout the remnant interior (i.e., $\overline{v}_{\rm IME}(x')={\rm constant}$, irrespective of $x'$).
If the ejecta is somehow decelerated, $\overline{v}_{\rm IME}$ should become lower and not follow the constant trend.
Therefore $\overline{v}_{\rm IME}$ is  not equivalent to the velocity itself  but should be considered as an indicator of the deceleration, i.e., ``deviation'' from expected value assuming only a high-velocity expansion.
With this property in mind, we present profiles  of calculated $\overline{v}_{\rm IME}$ in Figure~\ref{fig:v_IME}.
Interestingly,  $\overline{v}_{\rm IME}$ keeps constant within the radius of $\sim200\arcsec$ above which it significantly drops from $\sim4000$~km~s$^{-1}$ to $\sim1000$~km~s$^{-1}$ in almost every direction.  
Although the parameter $\overline{v}_{\rm IME}$ does not directly mean  a decelerated velocity as noted above, it is obvious that the velocity structure evidently contradict the expectation from the uniform expansion.
We thus conclude that the ejecta was decelerated in recent past by dense gas in the vicinity of Tycho's SNR.

\begin{figure}[ht]
\centering
\includegraphics[width=8.4cm]{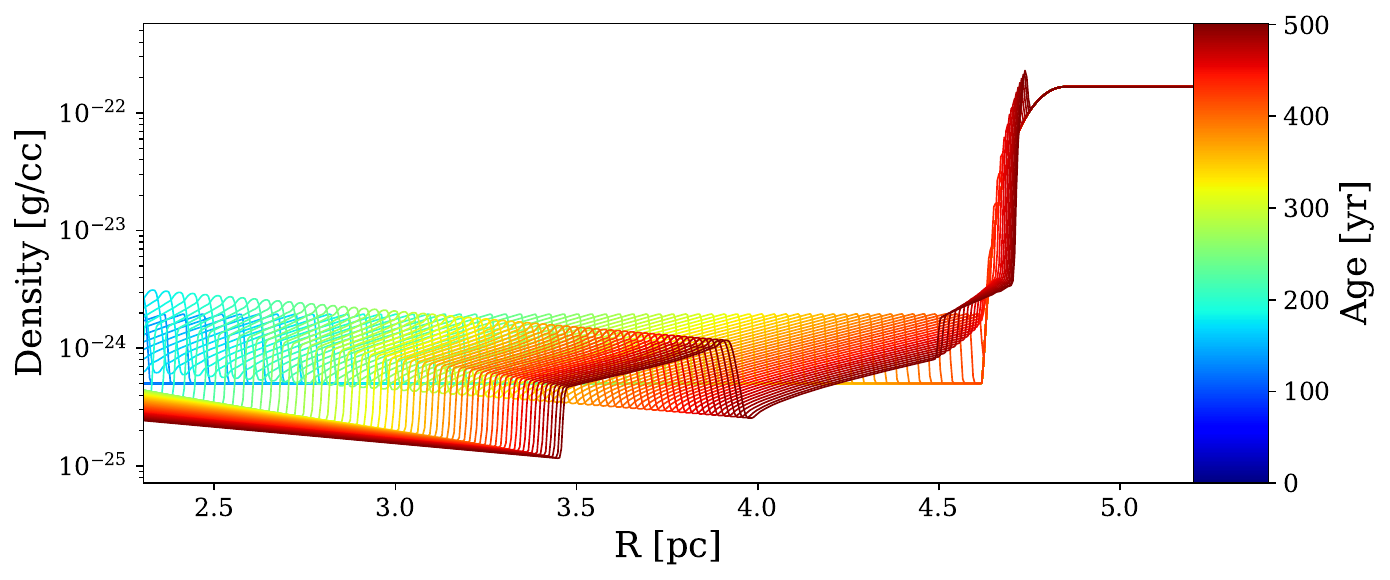}
\includegraphics[width=8.4cm]{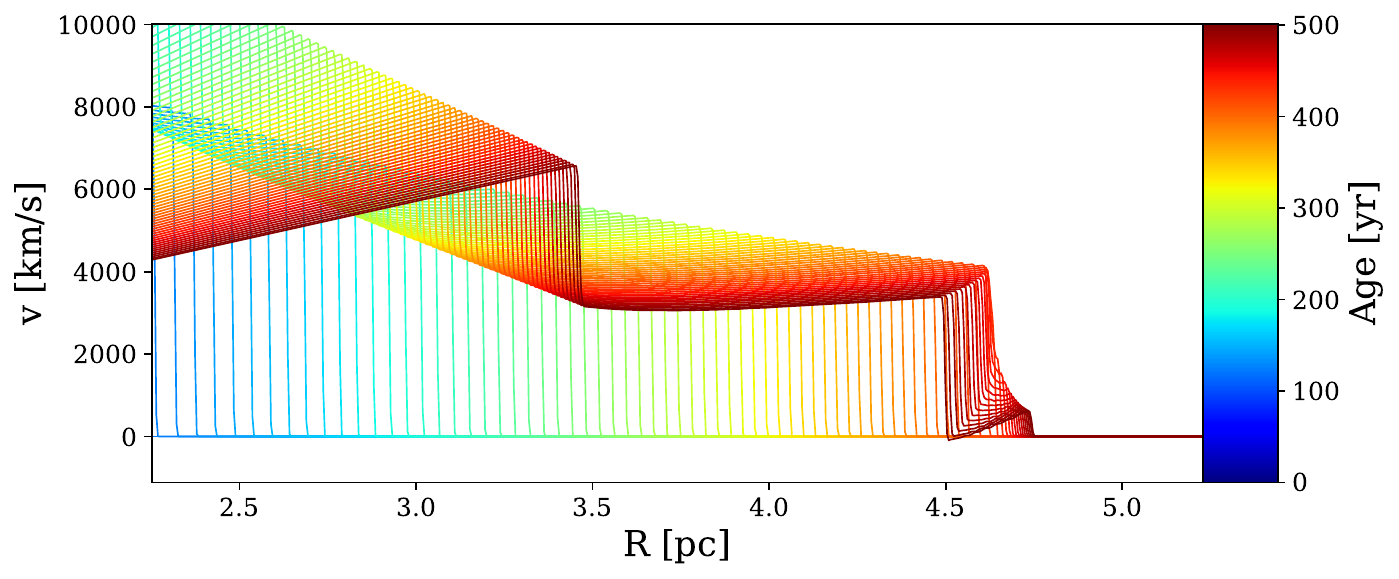}
\includegraphics[width=8.4cm]{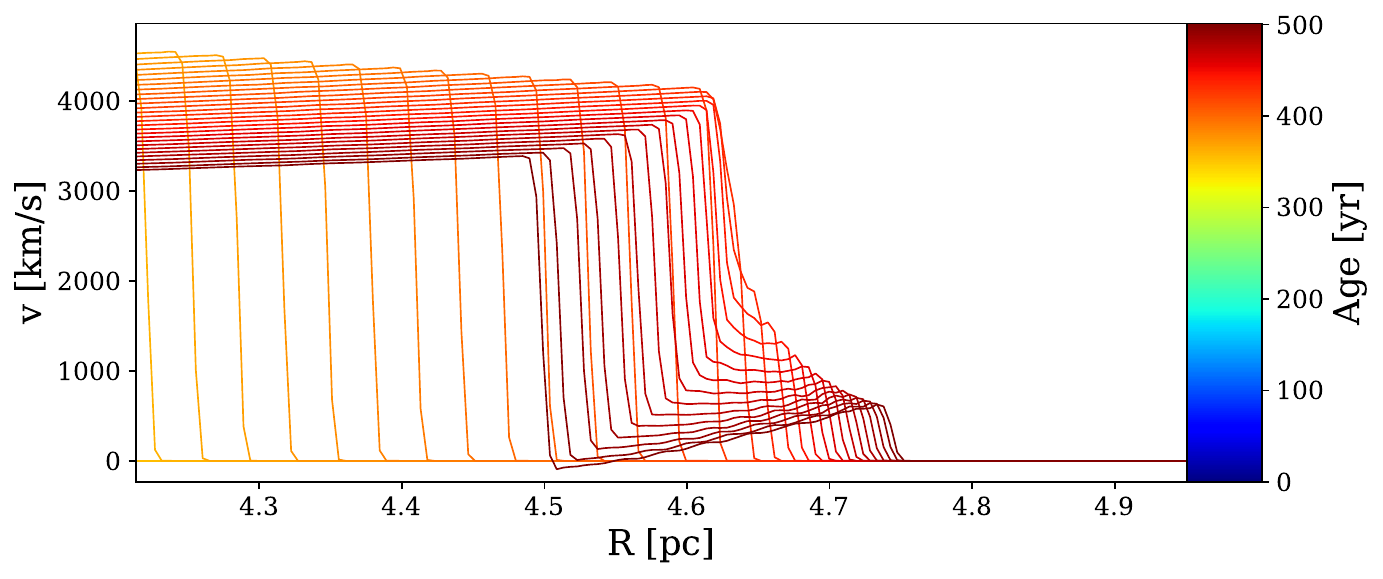}
\caption{Results of a hydrodynamical simulation. Top and middle panels display the density and velocity profiles with radius, respectively. The bottom panel shows the zoom-in view of the velocity profile near the outer edge, where the reflection shock is generated.  }
\label{fig:simulation}
\end{figure}

When an expanding SNR shock collides with a dense material, numerical simulations suggest that  a ``reflection (reflected) shock'' traverses back into the ejecta \citep[cf.][]{Dwarkadas05, Orlando22}.
Several lines of observational evidence are also reported from core-collapse SNRs interacting with molecular clouds \citep[e.g.,][]{Inoue12, Sato18}.
In the case of Tycho's SNR, as the previous works suggest \citep[e.g.,][]{Tanaka21, Kobashi23}, it is more reasonable to expect a cavity wall formed by the progenitor system before the SN explosion.
Assuming that the cavity had a radius of 4.5~pc with a density of 0.3~cm$^{-3}$ \citep[see also][]{Zhou16}, we present a hydrodynamical calculation of shock propagation as shown in Figure~\ref{fig:simulation}.
In our setup, a blast wave generated by a normal Type Ia SN is colliding with a wall with a number density of 100~cm$^{-3}$, which is consistent with the estimate  given by \citet{Tanaka21}.
The result demonstrates that the blast wave  hits  the wall and is decelerated rapidly, which gives a discontinuous low-velocity structure near the edge.
We confirmed that the velocity of the decelerated region is typically $\sim1000$~km~s$^{-1}$ during $\sim20$~yr after the collision.

As our simulation predicts, the generated reflection shock will move backward and converge onto the reverse shock within a few tens of years.
It provides important constraints on when (and where) Tycho's SNR reached the cavity wall.
Notably this timescale is fairly in agreement with the previous proper-motion measurement \citep{Tanaka21}, in which the authors observed a shock deceleration during last $\sim15$~yr in the southwest.
We  also infer that the cavity was nearly spherically symmetric around the SN explosion.  
We further found that the reflection shock forms a positive velocity gradient to the outside ($R=4.5$--4.7~pc in Figure~\ref{fig:simulation}); the trend is suggestively similar to the $\overline{v}_{\rm IME}$ profiles shown in Figure~\ref{fig:v_IME} (for example, $\theta=180^{\circ}$--$225^{\circ}$).
We thus conclude that our result provides further evidence of the cavity formed by a wind from the white dwarf during mass accretion  in the SD progenitor system  \citep{Hachisu96}.

Although there remains large uncertainties in the measurement with the CCDs, future observations with higher energy resolution will provide a more detailed velocity structure, which hints at the environment of Tycho's SNR and its progenitor system.
Our method with high angular resolution microcalorimeters such as Athena \citep{Barret18} and  Lynx \citep{Gaskin19} will provide a new way to distinguish between the SD and DD progenitor scenarios for Type Ia SNRs.

\section{Conclusions} \label{sec:con}
We revealed a global three-dimensional velocity structure of Tycho's SNR using XMM-Newton data, and investigated the presence of a wind bubble which was suggested by a recent proper-motion observation \citep{Tanaka21}.
We performed a spatially resolved  spectroscopy of the remnant (845 grid cells in total) and found that emission lines of the ejecta such as Si He$\alpha$ can be explained by a blue/red-shifted thermal model in almost all regions. 
While the obtained velocity structure approximately indicates a spherically symmetric expansion, the radial profile drops down far inside the edge in almost all directions, which implies a significant deceleration of blast waves around the remnant.

We converted the obtained line-of-sight velocities $v_{\parallel}$ to expansion velocities $v_{\rm IME}$ and found possible evidence that  ejecta just behind the CD was slowed down  to $v_{\rm IME}\sim1000$~km~s$^{-1}$, whereas global expansion velocity is $v_{\rm IME}=3870\pm40$~km~s$^{-1}$.
We also confirmed that average radii of RS and ejecta are $r_{\rm{RS}}=175\arcsec\pm3\arcsec$ and $r_{\rm{ejecta}}=210\arcsec\pm1\arcsec$, respectively, which are in good agreement with other estimations.
Our result strongly suggests a recent collision of Tycho's SNR with dense surrounding materials, supporting the presence of the cavity wall (wind bubble).

We performed a hydrodynamical calculation of shock propagation into a dense material.
The result indicates that a reflection shock is formed behind the forward shock and causes a significant deceleration, which can consistently explain the observed velocity structure.
We thus conclude that our result supports the presence of the cavity and a dense wall, and therefore the SD scenario is preferable for the origin of Tycho's SNR.
Since our study is based on the CCD data, a further analysis with a better energy resolution with microcalorimeters will reveal more detailed velocity structure of this remnant.
Also note that the method we present here is applicable to other Type Ia SNRs if we use future observatories like Athena and Lynx.
In addition to ways to test explosion models from SNR morphology \citep{Ferrand21} or abundances \citep{Yamaguchi15}, the velocity structure will provide useful information on the presence of stellar wind bubbles and hence will be helpful to  identify the SD or DD progenitor scenario.

\begin{acknowledgments}

T.K. appreciates all supports for his short staying at Kyoto by Takeshi Go Tsuru, Toshihiro Fujii, and their group members at Kyoto University, and Hein Mallee and Takako Okamoto at the Research Institute for Humanity and Nature. 
The authors are very grateful to Paul Plucinsky, Hirokazu Odaka, Shigehiro Nagataki, and Gilles Ferrand for giving us precious advice on this study.
The authors also thank Tsuyoshi Inoue for the helpful discussion on the shock physics.
This work is mainly supported by the Japan Society for the Promotion of Science (JSPS) KAKENHI Grant Nos. JP21J11443 (T.K.), JP22H01265 (H.U.), JP20H00174 (K.M.), JP19H01936 (T.T.), JP21H04493 (T.T.), and JP19K03908 (A.B.).

\end{acknowledgments}

\begin{appendix}\label{sec:app}
\section{Sacrificial charge issue encountered in analysis of Chandra data}\label{sec:app_chandra}
In this appendix, we briefly present an effect of ``sacrificial charge'' that we encountered in our  analysis of Chandra ACIS data.
This issue will become  severe when a bright diffuse source is observed with a CCD detector.
Since  a signal charge is lost during readout with a certain probability due to traps of the lattice of silicon, an enhancement of  charge transfer inefficiency (CTI) cannot  be generally ignored for  in-orbit photon-counting CCDs \citep{Townsley00}.
In order to recover energy losses, the CTI correction method is therefore applied  in a standard  reduction and analysis of Chandra  ACIS dataset \citep{Plucinsky02}.
However, if a very high-count-rate X-ray object is observed with the ACIS CCDs, charge deposited along the readout path  fills the traps during readout and thus the CTI is unintentionally  and excessively corrected.
This is called sacrificial charge, which gives a wrong energy peak and line width \citep{Grant03}.
The  sacrificial charge issue would become more  considerable for high-angular resolution instruments.

\begin{figure}[ht]
\centering
\includegraphics[width=13cm]{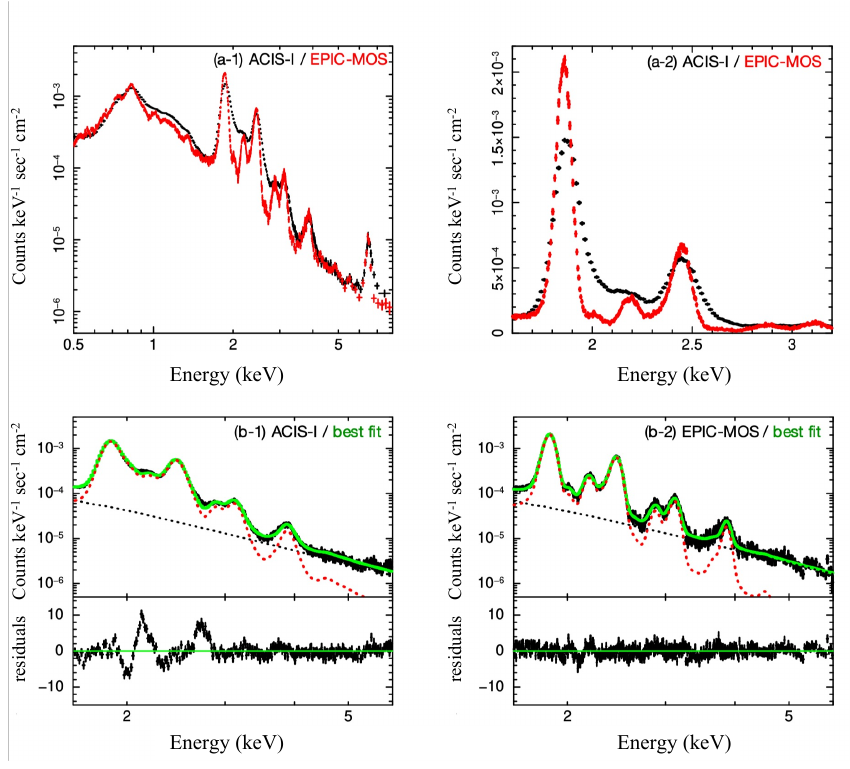}
\caption{Spectra of a region in Tycho's SNR obtained with the XMM-Newton MOS (red) and the Chandra ACIS (black). 
The top left and right panels show the same spectra but x-axis is in the logarithmic and linear scales, respectively.
The bottom panels show the ACIS (left) and MOS (right) spectra displayed with the black data points.
The best-fit model obtained with the MOS data is overlaid onto each spectrum; the black, red, and green lines indicate the power-law, thermal, and total model shapes, respectively.}
\label{fig:chandra_xmm}
\end{figure}

Figure~\ref{fig:chandra_xmm} shows a comparison between spectra of Tycho's SNR, for which the same region is extracted  with XMM-Newton and Chandra.
We found  a clear difference between the two spectra and confirmed that the ACIS spectrum shows a broader and more blue-shifted feature, even if taking into account the difference in energy resolution between each other.
Most of the ACIS spectra are suffering from this problem; estimated expansion velocities exceed 3000~km~s$^{-1}$ even around the edge of the remnant.
These results are clearly attributed to the effect of the sacrificial charge explained above.
We  therefore did not use the Chandra data for our analysis.

\section{Modeling of line-of-sight velocity structure}\label{sec:los}
As shown in Figure~\ref{fig:tycho_schematics}, we assume that Tycho's SNR is spherically symmetric \citep{Hayato10} and the X-ray emitting ejecta (IME) is filled between the RS and the CD with an expansion velocity $v_{\rm IME}$.
In a two-dimensional plane, which is a cut surface  of the angular direction of the remnant, we define the $x$ and $z$ axes as the angular and the line-of-sight directions, respectively.
In any position $x=x'$ and $z=z'$ ($r' \equiv \sqrt{ x'^2 + z'^2}$), the line-of-sight velocity is described as $v(x'; z')=v_{\rm IME} \cos{\varphi}$, where $\varphi$ is the angle between the direction of $v_{\rm IME}$ and the $z$ axis: $\rm{tan}\varphi=x'/z'$.
Here, an \textit{integrated} line-of-sight velocity $f(x';R)$ at $x=x'$ is calculated as
\begin{equation}
f(x';R) = \int_{z=0}^{z'} {\rm d} z \ v_{\rm IME}\cos{\varphi},
\end{equation}
where $R$ is the radius of the circumference of the ejecta on the  two-dimensional plane.
By defining  $\Theta \equiv \arcsin{(x'/R)}$, we can modify this equation as
\begin{eqnarray}
f(x';R)  &=& \int_{\varphi=\pi/2}^{\Theta} {\rm d}\varphi \ \frac{{\rm d}}{{\rm d}\varphi} \left( \frac{x'}{\tan{\varphi}} \right) v_{\rm IME} \cos{\varphi} \\ 
&=& \int_{\varphi=\pi/2}^{\Theta} {\rm d}\varphi \ \left( -\frac{x'}{\sin^2{\varphi}} \right) v_{\rm IME} \cos{\varphi} \\
&=& v_{\rm IME}x' \left[ \frac{1}{\sin{\varphi}} \right]_{\varphi=\pi/2}^{\Theta}  \\ 
&=&v_{\rm IME}(R-x').
\end{eqnarray}
The observed  line-of-sight velocity $F_1(x'; R)$ should be the density-averaged value of $f(x';R)$.
If the ejecta density is uniform, $F_1(x'; R)$ is calculated as 
\begin{equation}
F_1(x'; R) = \frac{f(x';R)}{z'^2} = \frac{f(x';R)}{\sqrt{R^2-x'^2}} = v_{\rm IME}\sqrt{\frac{R-x'}{R+x'}}.
\end{equation}
On the other hand, if the ejecta is filled between  $R=R_1$ and $R=R_2$ ($R_1 < R_2$), the line-of-sight velocity $F_2(x'; R)$ can be described as 
\begin{eqnarray}
F_2(x';R_1,R_2) &=& \frac{f(x';R_2)-f(x';R_1)}{\sqrt{R_2^2-x'^2}-\sqrt{R_1^2-x'^2}} \\ 
&=& v_{\rm IME} \frac{\sqrt{R_1^2-x'^2}+\sqrt{R_2^2-x'^2}}{R_1+R_2} \\ 
&=& v_{\rm IME} \left( \frac{R_1}{R_1+R_2} \sqrt{1 - \left( \frac{x'}{R_1} \right)^2} + \frac{R_2}{R_1+R_2} \sqrt{1 - \left( \frac{x'}{R_2} \right)^2} \right).
\end{eqnarray}
Here we assume that the ejecta is  filled between the RS and an outer boundary, hence $R_1=r_{\rm RS}$, $R_2=r_{\rm ejecta}$, and $F_1(x'; R)=F_2(x';R_1,R_2)$ at $x'=r_{\rm RS}$. 
Under this assumption, we derive the following equations for the line-of-sight expansion velocity $v_{\parallel}$ at $x=x'$:
\begin{equation}
v_{\parallel}(x') =  \\
\left\{
\begin{array}{ll}
v_{\rm IME} \left( \frac{r_{\rm RS}}{r_{\rm RS}+r_{\rm ejecta}} \sqrt{1 - \left( \frac{x'}{r_{\rm RS}} \right)^2}+ \frac{r_{\rm ejecta}}{r_{\rm RS}+r_{\rm ejecta}} \sqrt{1 - \left( \frac{x'}{r_{\rm ejecta}} \right)^2} \right) &  (0 \leq x' < r_{\rm{RS}}), \\
v_{\rm IME} \sqrt{\frac{r_{\rm ejecta}-x'}{r_{\rm ejecta}+x'}} & (r_{\rm{RS}} \leq x' < r_{\rm{ejecta}}), \\
0 &  (r_{\rm{ejecta}} \leq x').\\
\end{array}
\right.
\end{equation}

\end{appendix}

\facilities{{\it XMM-Newton} (MOS) \citep{Turner01}}

\software{SAS 19.1.0 \citep{Gabriel04}}


\bibliography{bibliography}{}
\bibliographystyle{aasjournal}

\end{document}